\documentclass[11pt, a4paper]{article}

\usepackage{float}

\usepackage{times}

\usepackage{graphicx}

\usepackage{url}
\usepackage{amsmath}
\usepackage{amssymb}
\usepackage{cite}

\usepackage{pbox}
\usepackage{booktabs}
\usepackage[table]{xcolor}

\usepackage{tikz}
\usetikzlibrary{arrows, petri, topaths}
\usepackage{tkz-berge}
\usepackage[utf8]{inputenc}

\title{Who can replace Xavi? A passing motif analysis of football players.}
\author{Javier L\'opez Pe\~{n}a \footnote{Kickdex Ltd, and Department of Mathematics, University College London. \texttt{javier@kickdex.com}} \and Ra\'ul S\'anchez Navarro \footnote{Kickdex Ltd, and Universidad de Granada. \texttt{snchznvrr@gmail.com}}}

\date{}

\begin{document} \thispagestyle{empty}

\twocolumn
\maketitle

\thispagestyle{empty}
\begin{abstract}
    Traditionally, most of football statistical and media coverage has been focused almost exclusively on goals and (ocassionally) shots. However, most of the duration of a football game is spent away from the boxes, passing the ball around.
    The way teams pass the ball around is the most characteristic measurement of what a team's ``\emph{unique style}'' is.
    In the present work we analyse passing sequences at the player level, using the different passing frequencies as a ``digital fingerprint'' of a player's style. The resulting numbers provide an adequate feature set which can be used in order to construct a measure of similarity between players. Armed with such a similarity tool, one can try to answer the question: `\emph{Who might possibly replace Xavi at FC Barcelona?}'
\end{abstract}

\section{Introduction}

Association football (simply referred to as \emph{football} in the forthcoming) is arguably the most popular sport in the world.
Traditionally, plenty of attention has been devoted to goals and their distribution as the main focus of football statistics. However, shots remain a rare occurrence in football games, to a much larger extent than in other team sports.

Long possessions and paucity of scoring opportunities are defining features of football games. Passes, on the other hand, are two orders of magnitude more frequent than goals, and therfore constitute a much more appropriate event to look at when trying to describe the elusive quality of `\emph{playing style}'. Some studies on passing have been performed, either at the level of passing sequences distributions (cf \cite{Hughes2005, LopezUNa, Reep1968}), by studying passing networks \cite{Duch2010, Lopez2013, Cotta2011}, or from a dynamic perspective studying game flow \cite{Brillinger2007}, or passing \emph{flow motifs} at the team level \cite{GyarmatiUN}, where passing flow motifs (developed following \cite{Milo2002}) were satisfactorily proven by Gyarmati, Kwak and Rodríguez to set appart passing style from football teams from randomized networks.

In the present work we ellaborate on \cite{GyarmatiUN} by extending the flow motif analysis to a player level. We start by breaking down all possible 3-passes motifs into all the different variations resulting from labelling a distinguished node in the motif, resulting on a total of 15 different 3-passes motifs at the player level (stemming from the 5 motifs for teams). For each player in our dataset, and each game they partitipate in, we compute the number of instances each pattern occurs. The resulting 15-dimensional distribution is used as a fingerprint for the player style, which characterizes what type of involvement the player has with his teammates.

The resulting feature vectors are then used in order to provide a notion of similarity between different football players, providing us with a quantifiable measure on how \emph{close} the playing styles between any two arbitrary players are. This is done in two different ways, first by performing a Clustering Analysis (with automatic cluster detection) on the feature vectors, which allow us to identify 37 separate groups of similar players, and secondly by defining a distance function (based on the mean features z-scores) which consequently is used to construct the distance similarity score.

As an illustrative example, we perform a detailed analysis of all the defined quantities for Xavi Hernández, captain of FC Barcelona who just left the team after many years in which he has been considered the flagship of the famous \emph{tiki-taka} style both for his club and for the Spanish national team. Using our data-based \emph{style fingerprint} we try to address the pressing question: which player could possibly replace the best passer in the world?

\section{Methodology}

The basis of our analysis is the study of passing subsequences. The passing style of a team is partially encoded, from an static point of view, in the passing network (cf. \cite{Lopez2013}). A more dynamical approach is taken in \cite{GyarmatiUN}, where passing subsequences are classified (at the team level) through ``flow motifs'' of the passing network.

Inspired by the work on flow motifs for teams, we carry out a similar analysis at the player level. We focus on studying flow motifs corresponding to sequences of three consecutive passes. Passing motifs are not concerned with the names of the players involved on a sequence of passes, but rather on the structure of the sequence itself. From a team's point of view, there are five possible variations: ABAB, ABAC, ABCA, ABCB, and ABCD (where each letter represents a different player within the sequence).

\begin{figure}[ht]

    \begin{tikzpicture}[scale=0.8,transform shape, >=stealth']
        \Vertex[x=0, y=0]{A}
        \Vertex[x=2, y=0]{B}
        \tikzset{EdgeStyle/.style={->}}
        \Edge[style={bend left = 15}](A)(B)
        \Edge(B)(A)
        \Edge[style={bend right = 15}](A)(B)
    \end{tikzpicture} \quad
    \begin{tikzpicture}[scale=0.8,transform shape, >=stealth']
        \Vertex[x=0, y=0]{A}
        \Vertex[x=1, y=1]{B}
        \Vertex[x=2, y=0]{C}

        \tikzset{EdgeStyle/.style={->}}
        \Edge[style={bend left = 15}](A)(B)
        \Edge[style={bend left = 15}](B)(A)
        \Edge[style={bend right = 15}](A)(C)
    \end{tikzpicture} \quad
    \begin{tikzpicture}[scale=0.8,transform shape, >=stealth']
        \Vertex[x=0, y=0]{A}
        \Vertex[x=0.75, y=1]{B}
        \Vertex[x=1.5, y=0]{C}

        \tikzset{EdgeStyle/.style={->}}
        \Edge[style={bend left = 15}](A)(B)
        \Edge[style={bend left = 15}](B)(C)
        \Edge[style={bend left = 15}](C)(A)
    \end{tikzpicture}
    \\[10pt]
    \centering
    \begin{tikzpicture}[scale=0.8,transform shape, >=stealth']
        \Vertex[x=0, y=0]{A}
        \Vertex[x=1.5, y=0]{B}
        \Vertex[x=3, y=0]{C}

        \tikzset{EdgeStyle/.style={->}}
        \Edge(A)(B)
        \Edge[style={bend left = 15}](B)(C)
        \Edge[style={bend left = 15}](C)(B)
    \end{tikzpicture} \qquad
    \begin{tikzpicture}[scale=0.8,transform shape, >=stealth']
        \Vertex[x=0, y=0]{A}
        \Vertex[x=1, y=0]{B}
        \Vertex[x=2, y=0]{C}
        \Vertex[x=3,y=0]{D}
        \tikzset{EdgeStyle/.style={->}}
        \Edge(A)(B)
        \Edge(B)(C)
        \Edge(C)(D)
    \end{tikzpicture}
    \caption{The five team flow motifs}
\end{figure}
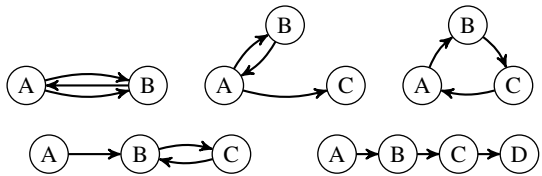

The situation is different when looking at flow motifs from an specific player's point of view, as that player needs to be singled out within each passing sequence. Allowing for variation of a single player's relative position within a passing sequence, the total numer of motifs increases to fifteen. These patterns can all be obtained by swapping the position of player A with each of the other players (and relabelling if necessary) in each of the five motifs for teams. Adopting the convention that our singled-out player is always denoted by letter `A', the resulting motifs can be labelled as follows (the basic team motif shown in bold letters):

\begin{table}[h]
\centering
\begin{tabular}{c}
    \textbf{ABAB}, BABA \\
    \textbf{ABAC}, BABC, BCBA \\
    \textbf{ABCA}, BACB, BCAB \\
    \textbf{ABCB}, BACA, BCAC \\
    \textbf{ABCD}, BACD, BCAD, BCDA

\end{tabular}
\end{table}

When tracking passing sequences, we will consider only possessions consisting of uninterrupted consecutive events during which the ball is kept under control by the same team. As such, we will consider than a possession ends any time the game gets interrupted or an action does not have a clear passing target. In particular, we will consider that posessions get interrupted by fouls, by the ball getting out of play, whenever there is a ``divided ball'' (eg an aerial duel), by clearances, interceptions, passes towards an open space without a clear target, or by shots, regardless on who gets to keep the ball afterwards. The motivation for this choice is that we are trying to keep track of game style through controlled, conscious actions. It is worth noting that here we are using a different methodology from the one in \cite{GyarmatiUN} (where passes are considered to belong to the same sequence if they are separated by less than five seconds).

Our analysed data consists of all English Premier League games over the last five seasons (comprising a total of 1900 games and 1402195 passes), all Spanish Liga games over the last three season (1140 games and 792829 passes), and the last season of Champions League data (124 games and 105993 passes). To reduce the impact of outliers, we have limited our study to players that have participated in at least 19 games (half a season). In particular, this means that only players playing the English and Spanish leagues are tracked in our analysis. Unfortunately, at the time of writing we do not have at our dispossal enough data about other European big leagues to make the study more comprehensive.

The resulting dataset contains a total of 1296 players. For each of the analyzed players, we compute the average number of occurrences of each of the fifteen passing motifs listed above, and use the results as the features vector in order to describe the player's style. For some of the analysis which require making different types of subsequences comparable, we replace the feature vector by the corresponding z-scores (where for each feature mean and standard deviation are computed over all the players included in the study).

Our analysis uses raw data for game events provided by Opta.
Data munging, model fitting, analysis, and chart plotting were performed using IPython \cite{ipython} and the python scientific stack \cite{numpy, matplotlib}.

\section{Analysis and results}

\subsection*{Summary statistics and motifs distributions}

\begin{figure*}[h]
    \centering
    \includegraphics[width=\textwidth]{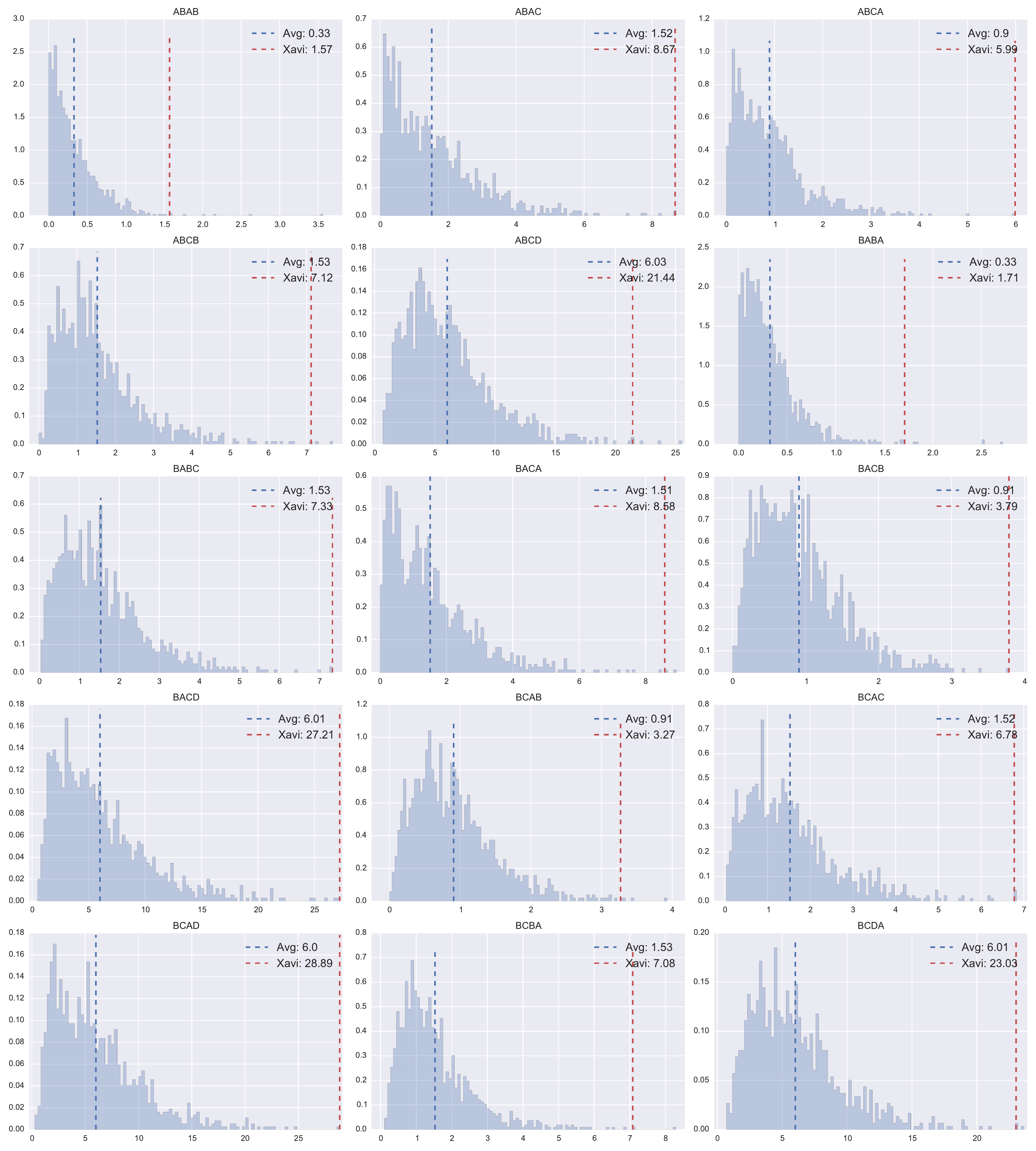}
    \caption{Passign motifs distributions}
    \label{fig:motif_dists}
\end{figure*}

A summary analysis of the passing motifs is shown in Table \ref{table:summary}. Perhaps unsurprisingly,
the maximum value for almost every single motif is reached by a player from FC Barcelona, the only exception being Yaya Touré.\footnote{Touré \textbf{did} play for FC Barcelona, however, our dataset only contains games in which he played for Manchester City. On the opposite side, we only have data for Thiago Alcántara as a Barcelona player as our dataset does not include the German Bundesliga.} Figure \ref{fig:motif_dists} shows the frequency distributions for player values at every kind of motif, and the relative position of Xavi within those distributions.

\begin{table}[h]
\begin{tabular}{lrrrl}
    \toprule
    \textbf{Motif} &  Mean &   Std &  Max &            Player \\
    \midrule
    \textbf{ABAB}  &  0.33 &  0.31 &     3.56 &        Dani Alves \\
    \textbf{ABAC}  &  1.52 &  1.30 &     8.71 &  Thiago Alcántara \\
    \textbf{ABCA}  &  0.90 &  0.73 &     5.99 &              Xavi \\
    \textbf{ABCB}  &  1.53 &  1.08 &     7.69 &   Sergio Busquets \\
    \textbf{ABCD}  &  6.03 &  3.62 &    25.53 &        Jordi Alba \\
    \textbf{BABA}  &  0.33 &  0.29 &     2.72 &      Lionel Messi \\
    \textbf{BABC}  &  1.53 &  1.07 &     7.33 &              Xavi \\
    \textbf{BACA}  &  1.51 &  1.28 &     8.94 &  Thiago Alcántara \\
    \textbf{BACB}  &  0.91 &  0.59 &     3.79 &              Xavi \\
    \textbf{BACD}  &  6.01 &  4.17 &    27.21 &              Xavi \\
    \textbf{BCAB}  &  0.91 &  0.58 &     3.93 &        Yaya Touré \\
    \textbf{BCAC}  &  1.52 &  1.08 &     6.83 &        Jordi Alba \\
    \textbf{BCAD}  &  6.00 &  4.11 &    28.89 &              Xavi \\
    \textbf{BCBA}  &  1.53 &  1.03 &     8.29 &   Sergio Busquets \\
    \textbf{BCDA}  &  6.01 &  3.47 &    23.64 &        Dani Alves \\
    \bottomrule
\end{tabular}
\caption{Motif average values and players with highest values}
\label{table:summary}
\end{table}

We can see how Xavi dominates the passing, being the player featuring the highest numbers in five out of the fifteen motifs. Table \ref{table:xavi_vals} shows all the values and z-scores for Xavi. It is indeed remarkable that he manages to be consistently over four standard deviations away from the average passing patters, and particularly striking his astonishing z-score of 6.95 in the \textbf{ABCA} motif, which corresponds to being the starting and finishing node of a triangulation. To put this number in context, if we were talking about random daily events, one would expect to observe such a strong deviation from the average approximately once every billion years!\footnote{From a very rigorous point of view, actual passing patterns are neither random nor normally distributed. Statistical technicalities notwithstanding, Xavi's z-scores are truly off the charts!}

\begin{table}[h]
\centering
\begin{tabular}{lrr}
    \toprule
    \textbf{Motif} &   Value &  z-score \\
    \midrule
    \textbf{ABAB}  &   1.57 &     3.97 \\
    \textbf{ABAC}  &   8.67 &     5.49 \\
    \textbf{ABCA}  &   5.99 &     6.95 \\
    \textbf{ABCB}  &   7.12 &     5.19 \\
    \textbf{ABCD}  &  21.44 &     4.26 \\
    \textbf{BABA}  &   1.71 &     4.71 \\
    \textbf{BABC}  &   7.33 &     5.41 \\
    \textbf{BACA}  &   8.58 &     5.51 \\
    \textbf{BACB}  &   3.79 &     4.88 \\
    \textbf{BACD}  &  27.21 &     5.08 \\
    \textbf{BCAB}  &   3.27 &     4.06 \\
    \textbf{BCAC}  &   6.78 &     4.86 \\
    \textbf{BCAD}  &  28.89 &     5.57 \\
    \textbf{BCBA}  &   7.08 &     5.40 \\
    \textbf{BCDA}  &  23.03 &     4.90 \\
    \bottomrule
\end{tabular}
\caption{Motif values and z-scores for Xavi}
\label{table:xavi_vals}
\end{table}

\subsection*{Clustering and PCA}

\begin{figure}[h]
    \centering
    \includegraphics[width=\columnwidth]{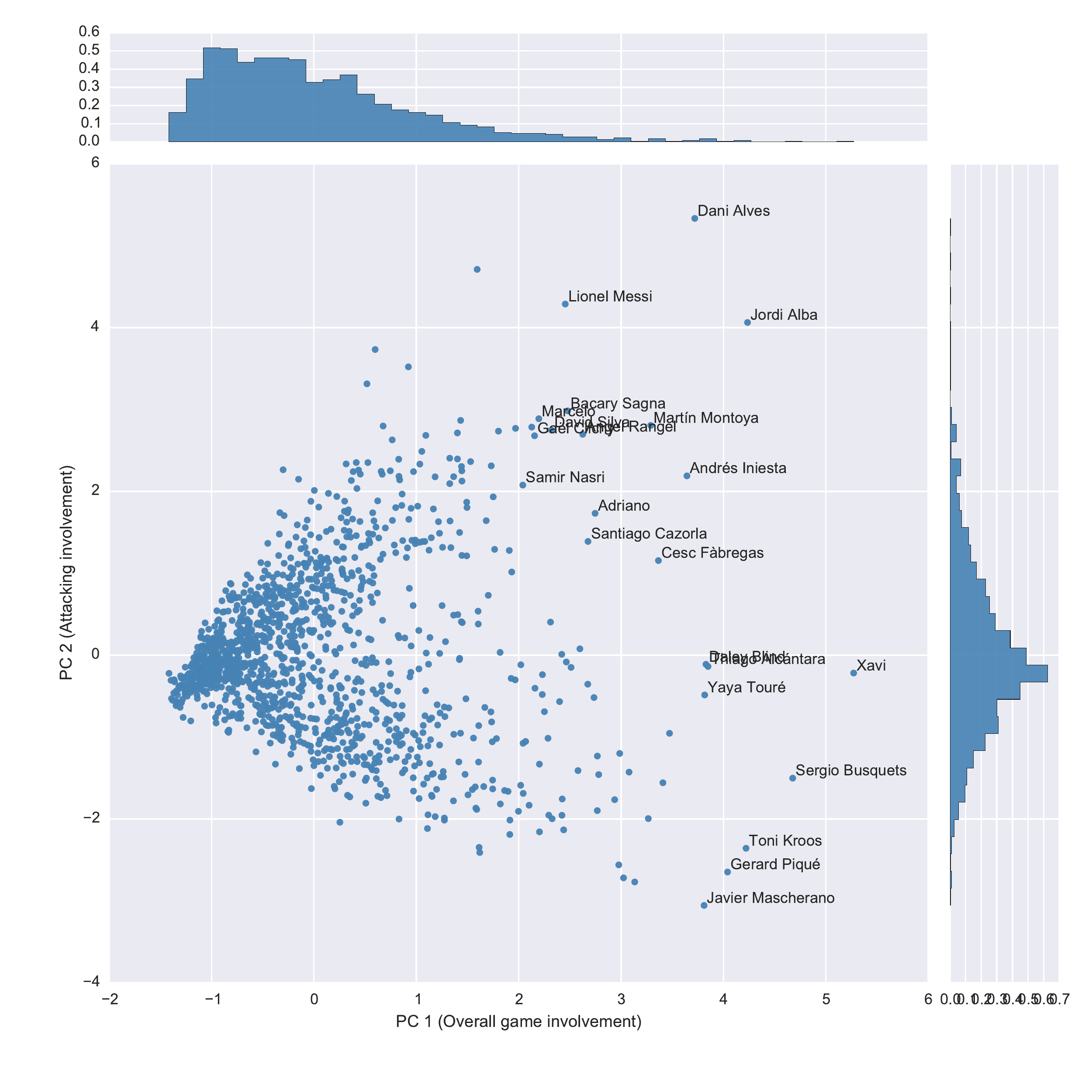}
    \caption{Pricipal Component Analysis, with labels for small AP clusters}
    \label{fig:pca_plot_hists}
\end{figure}

Using the passing motifs means as feature vectors, we performed some clustering analysis on our player set.
The Affinity Propagation method with a damping coefficient of 0.9 yields a total of 37 clusters with varying number of players, listed in Table \ref{table:ap_clusters}, where a representative player for every cluster is also listed. The explicit composition of each of the clusters of size smaller than 10 is shown in Table \ref{table:small_clusters}.
Once again we can observe how the passing style of Xavi is different enough from everyone else's to the extent that he gets assignated to a cluster of his own!

Figure \ref{fig:pca_plot_hists} shows the relative players feature vectors, plotted using the first two components of a Principal Component Analysis (after using a whitening transformation to eliminate correlation). The PC's coefficients, together with their explained variance ratio, are listed in Table \ref{table:pca}. After looking at Figure \ref{fig:pca_plot_hists}, one can think of the first principal component (PC 1) as a measurement of overall involvement on the game, whereas the second principal componen (PC 2) separates players depending on their positional involvement, with high positive values highlight players playing on the wings and with a strong attacking involvement, and smaller values relate to a more purely defensive involvement. Special mention on this respect goes to Dani Alves and Jordi Alba, who in spite of playing as fullbacks display a passing distribution more similar to the ones of forwards than to other fullbacks. The plot also shows how Xavi has the highest value for overall involvement and a balanced involvement between offensive and defensive passing patterns.

\begin{table}[h]
    \centering
    \begin{tabular}{lrr}
        \toprule
        {} &   \textbf{PC 1} &   \textbf{PC 2} \\
        \midrule
        \textbf{ABAB}  &  0.030 &  0.065 \\
        \textbf{ABAC}  &  0.153 & -0.019 \\
        \textbf{ABCA}  &  0.084 & -0.031 \\
        \textbf{ABCB}  &  0.127 & -0.091 \\
        \textbf{ABCD}  &  0.437 &  0.150 \\
        \textbf{BABA}  &  0.027 &  0.051 \\
        \textbf{BABC}  &  0.114 &  0.257 \\
        \textbf{BACA}  &  0.150 & -0.040 \\
        \textbf{BACB}  &  0.070 &  0.043 \\
        \textbf{BACD}  &  0.514 & -0.451 \\
        \textbf{BCAB}  &  0.064 &  0.086 \\
        \textbf{BCAC}  &  0.107 &  0.323 \\
        \textbf{BCAD}  &  0.511 & -0.310 \\
        \textbf{BCBA}  &  0.123 & -0.062 \\
        \textbf{BCDA}  &  0.406 &  0.690 \\
        \midrule
        \textbf{Explained variance} & 0.917   & 0.046 \\
        \bottomrule
    \end{tabular}
    \caption{Principal Components and their explained variance}
    \label{table:pca}
\end{table}

\subsection*{Player distance and similarity} 
\label{sub:player_distance_and_similarity}

Our feature vector can be used in order to define a measure of similarity between players.
Given a player $i$, let $\mathbf{v}_i$ denote the vector of z-scores in passing motifs for player $i$.
Our definition of \emph{distance} between two players $i$ and $j$ is simply the Euclidean distance between the corresponding (z-scores) feature vectors:
\[
    d(i, j) := \left\| \mathbf{v}_i - \mathbf{v}_j \right\|_2 = \sqrt{\sum_{m\in\text{motifs}} (v_{i,m} - v_{j,m})^2}
\]

This distance can be used as a measure of similarity between players, allowing us to establish how closely related are the passing patterns of any two given players. In more concrete terms, the coefficient of similarity is defined by
\[
    s(i, j) := \frac{1}{1 + d(i, j)}.
\]
This similarity score is always between 0 and 1, with 1 meaning that two players display an identical passing pattern.

The reason for choosing z-scores rather than raw values is to allow for a better comparison between different passing motifs, as using raw values would yield a distance dominated by the four motifs derived from \textbf{ABCD}, which show up in a frequency one order of magnitude higher than any other pattern. Table \ref{table:dists} shows a summary of the average and minimum distances for all the players in our dataset, showing that for an average player we can reasonably expect to find another one at a distance of $0.826 \pm 0.5$.

\begin{table}[ht]
    \centering
    \begin{tabular}{lrr}
\toprule
{} &  Mean &  Closest \\
\midrule
Avg value  &           4.471 &          0.826 \\
Std deviation  &           1.800 &          0.500 \\
Min value   &           3.188 &          0.178 \\
Max value  &          19.960 &          5.134 \\
\bottomrule
\end{tabular}
\caption{Average and closest player distances.}
\label{table:dists}
\end{table}

An immediate application of this is to find out, for a given player, who is his \emph{closest} peer, which will be the player displaying the most similar passing pattern. Table \ref{table:player_min_dists} shows the minimum distances to the ten bottom players (the ones with the smallest minimum distance, hence easier to replace) and the top 10 players (the ones with the hightest minimum distance, thus harder to replace). Once again, we can see how the top 10 players are dominated by FC Barcelona players.

\begin{table}[ht]
    \centering
    \begin{tabular}{lr|lr}
\toprule
           Player &  Closest  &            Player &  Closest \\
\midrule
  R Boakye &          0.18 &      A Rangel &          3.08 \\
           Tuncay &          0.18 &            Neymar &          3.26 \\
 J Arizmendi &          0.23 &        Y Touré &          3.92 \\
    J Roberts &          0.23 &  T Alcántara &          3.92 \\
  S Fletcher &          0.23 &    A Iniesta &          4.27 \\
     F Borini &          0.23 &   J Alba &          4.48 \\
   G Toquero &          0.24 &    D Alves &          4.48 \\
    Babá &          0.24 &        Xavi &          4.49 \\
 J Walters &          0.25 &      L Messi &          5.09 \\
   C Austin &          0.25 &   S Busquets &          5.13 \\
\bottomrule
\end{tabular}
\caption{Players minimum distance (bottom 10 and top 10)}
\label{table:player_min_dists}
\end{table}

Note that in some cases, the closest peer for a player happens to play for the same team, as it is the case for Jordi Alba, whose closest peer is Dani Alves. We decided against filtering closest player to search in team as it would make the analysis overly complicated due to constant player movement between teams.

Previous table shows that Xavi is amongst the hardest players to find a close replacement for. Table \ref{table:xavi} show the 20 players closest to Xavi. Among those, no one has a similarity score higher that 18.2\%, and only ten players have a score higher than 10\%.

\begin{table}[h]
    \centering
    \begin{tabular}{lrr}
        \toprule
        {Player}         &  Distance &  Similarity (\%) \\
        \midrule
        Yaya Touré       &     4.495 &      18.199 \\
        Thiago Alcántara &     5.835 &      14.631 \\
        Sergio Busquets  &     6.494 &      13.345 \\
        Andrés Iniesta   &     7.038 &      12.441 \\
        Cesc Fàbregas    &     7.377 &      11.938 \\
        Jordi Alba       &     7.396 &      11.910 \\
        Toni Kroos       &     7.853 &      11.296 \\
        Mikel Arteta     &     8.257 &      10.802 \\
        Michael Carrick  &     8.505 &      10.521 \\
        Santiago Cazorla &     8.515 &      10.509 \\
        Daley Blind      &     9.154 &       9.849 \\
        Paul Scholes     &     9.240 &       9.765 \\
        Gerard Piqué     &     9.524 &       9.502 \\
        David Silva      &     9.640 &       9.398 \\
        Marcos Rojo      &     9.671 &       9.371 \\
        Angel Rangel     &     9.675 &       9.368 \\
        Samir Nasri      &     9.683 &       9.360 \\
        Leon Britton     &     9.797 &       9.261 \\
        Aaron Ramsey     &     9.821 &       9.241 \\
        Martín Montoya   &     9.846 &       9.220 \\
        \bottomrule
    \end{tabular}
    \caption{Distances and similarity scores of the 20 players closest to Xavi.}
    \label{table:xavi}
\end{table}

\section{Conclusions and future work} 
\label{sec:conclusions}

We have shown how the flow motif analysis can be extended from teams to players. Although there is an added level of complexity raising from the increasing of the different motives, the resulting data does a good job classifying and discriminating players. Clustering analysis provides a reasonable grouping of players with similar characteristics, and the similarity score provides a quantifiable measure on how similar any two players are. We believe these tools can be useful for scouting and for early talent detection if implemented properly.

For future work, we plan to expand our dataset to cover all the major European leagues over a longer time span. A larger dataset would allow us to measure changes in style over a player's career, and perhaps to isolate a \emph{team factor} that would allow to estimate what would be a player's style if he were to switch teams. Another interesting thing to explore would be the density of each of the passing motifs according to pitch coordinates.

Coming back to our motivating question, who can replace Xavi at Barcelona? Amongst all the ten players that showing a similarity score bigger than 10, three are already at Barcelona (Busquets, Iniesta and Jordi Alba), and another three used to play there but left (Touré, Alcántara and Fàbregas). Arteta, Carrick and Cazorla are all in their thirties, ruling them out as a long-term replacement, and Toni Kroos plays for Barcelona arch-rivals Real Madrid, making a move quite complicated (although not impossible, as current Barcelona manager Luis Enrique knows very well), the only choices for Barcelona seem to be either to recover Alcántara or Fàbregas, or to reconvert Iniesta to play further away from the oposition box. A bolder move would be the Dutch rising star, Daley Blind (who used to play as a fullback, but has been tested as a midfielder over the last season in Van Gaal's Manchester United), hoping that the young could rise to the challenge.

Xavi's passing patter stands out in every single metric we have used for our analysis. Isolated in his own cluster, and very far away from any other player, all data seems to point out at the fact that Xavi Hernández is, literally, one of a kind.


\begin{table}[h]
    \centering
    \begin{tabular}{lr}
\toprule
         Representative Player &  Cluster size \\
\midrule
               Xavi &             1 \\
         Dani Alves &             2 \\
   Thiago Alcántara &             4 \\
        David Silva &             4 \\
       Gerard Piqué &             5 \\
       Bacary Sagna &             6 \\
               Isco &             8 \\
              Chico &            10 \\
   Mahamadou Diarra &            12 \\
        Jonny Evans &            12 \\
   Jordan Henderson &            15 \\
      Andreu Fontás &            17 \\
  Christian Eriksen &            17 \\
         Hugo Mallo &            18 \\
     Victor Wanyama &            19 \\
  César Azpilicueta &            19 \\
     Alberto Moreno &            20 \\
        Gareth Bale &            20 \\
          Fran Rico &            30 \\
       David de Gea &            36 \\
    Antolin Alcaraz &            36 \\
      Phil Jagielka &            39 \\
  Sebastian Larsson &            40 \\
     Liam Ridgewell &            41 \\
     Emmerson Boyce &            44 \\
               Nyom &            46 \\
         John Ruddy &            48 \\
       Adam Johnson &            52 \\
    Richmond Boakye &            57 \\
      Chechu Dorado &            61 \\
      Manuel Iturra &            62 \\
      Loukas Vyntra &            62 \\
      Kevin Gameiro &            72 \\
              Borja &            73 \\
       Rubén García &            85 \\
 Gabriel Agbonlahor &            90 \\
    Steven Fletcher &           113 \\
\bottomrule
\end{tabular}
\caption{Affinity propagation cluster sizes and representative players}
\label{table:ap_clusters}
\end{table}

\begin{table*}[h]
\centering
\begin{tabular}{ll}
\toprule
Size & Players  \\
\midrule
    1 & Xavi \\ \midrule
    2 & Dani Alves, Jordi Alba \\ \midrule
    4 & \pbox{15cm}{David Silva, Lionel Messi, \\
                    Samir Nasri, Santiago Cazorla} \\ \midrule
    4 & \pbox{15cm}{Andrés Iniesta, Cesc Fàbregas, \\
                    Thiago Alcántara, Yaya Touré} \\ \midrule
    5 & \pbox{15cm}{Daley Blind, Gerard Piqué, Javier Mascherano, \\
                    Sergio Busquets, Toni Kroos} \\ \midrule
    6 & \pbox{15cm}{Adriano, Angel Rangel, Bacary Sagna, \\
                    Gaël Clichy, Marcelo, Martín Montoya} \\ \midrule
    8 & \pbox{15cm}{Emre Can, Isco, James Rodríguez, Juan Mata, \\
                    Maicon, Mesut Özil, Michael Ballack, Ryan Mason} \\ \midrule
   10 & \pbox{15cm}{Ashley Williams, Carles Puyol, Chico, Marc Bartra, Marcos Rojo, \\
                    Michael Carrick, Mikel Arteta, Nemanja Matic, Paul Scholes, Sergio Ramos} \\ \midrule
   12 & \pbox{15cm}{Dejan Lovren, Garry Monk, John Terry, Jonny Evans, \\
                    Ki Sung-yueng, Matija Nastasic, Michael Essien, Morgan Schneiderlin, \\
                    Nabil Bentaleb, Per Mertesacker, Roberto Trashorras, Vincent Kompany} \\ \midrule
   12 & \pbox{15cm}{Aaron Ramsey, Alexandre Song, Fernandinho, Gareth Barry, \\
                    Jerome Boateng, Jonathan de Guzmán, Leon Britton, Luka Modric, \\
                    Mahamadou Diarra, Mamadou Sakho, Steven Gerrard, Xabi Alonso} \\ \midrule
   15 &  \pbox{15cm}{Ander Herrera, Eric Dier, Frank Lampard, Ivan Rakitic, \\
                     Jamie O'Hara, Jordan Henderson, Michael Krohn-Dehli, Rafael van der Vaart, \\
                     Rafinha, Sascha Riether, Scott Parker, Seydou Keita, \\
                     Steven Davis, Vassiriki Abou Diaby, Wayne Rooney} \\
\bottomrule
\end{tabular}
\caption{Affinity propagation clustering: Composition of small clusters}
\label{table:small_clusters}
\end{table*}

\end{document}